\documentclass{article}
\usepackage{amsmath, amssymb, amsthm,bbm,bm}
\usepackage{geometry}

\RequirePackage[numbers]{natbib}

\usepackage{orcidlink}

\theoremstyle{plain}

\newtheorem{rmk}{Remark}

\title{Bayesian Pliable Lasso with Horseshoe Prior for Interaction Effects in GLMs with Missing Responses
}
\author{The Tien Mai\orcidlink{0000-0002-3514-9636}}

\date{
\small
Norwegian Institute of Public Health, Oslo, 0456, Norway
\\
email: the.tien.mai@fhi.no
}

\begin{document}

\maketitle

\begin{abstract}
Sparse regression problems, where the goal is to identify a small set of relevant predictors, often require modeling not only main effects but also meaningful interactions through other variables. While the pliable lasso has emerged as a powerful frequentist tool for modeling such interactions under strong heredity constraints, it lacks a natural framework for uncertainty quantification and incorporation of prior knowledge. In this paper, we propose a Bayesian pliable lasso that extends this approach by placing sparsity-inducing priors, such as the horseshoe, on both main and interaction effects. The hierarchical prior structure enforces heredity constraints while adaptively shrinking irrelevant coefficients and allowing important effects to persist. 
We extend this framework to Generalized Linear Models (GLMs) and develop a tailored approach to handle missing responses. To facilitate posterior inference, we develop an efficient Gibbs sampling algorithm based on a reparameterization of the horseshoe prior. Our Bayesian framework yields sparse, interpretable interaction structures, and principled measures of uncertainty. Through simulations and real-data studies, we demonstrate its advantages over existing methods in recovering complex interaction patterns under both complete and incomplete data.

Our method is implemented in the package \texttt{hspliable} available on Github: \url{https://github.com/tienmt/hspliable}.
\end{abstract}


\section{Introduction}

Linear regression with high-dimensional data where the number of potential explanatory variables (predictors)  significantly exceeds the sample size, presents a fundamental challenge that transcends disciplines such as statistics and machine learning \citep{hastie2009elements,buhlmann2011statistics,giraud2021introduction,tibshirani2020pliable}.
In linear regression, it often requires identifying not only relevant main effects but also important interactions between predictors \citep{bien2013lasso,lim2015learning}. In many application areas such as genomics, neuroscience, and personalized medicine, interactions between predictors and modifying variables (e.g., environmental factors, patient characteristics) can reveal deeper insights into underlying mechanisms,
see for example \citep{wu2014integrative,wu2018identifying,hoogland2021tutorial,kim2021svreg,sun2024bhaft,d2025integrating}. 
However, the number of potential interaction terms increases rapidly with dimensionality, making it essential to use regularization techniques that promote sparsity while maintaining interpretability.

The pliable lasso, introduced by \cite{tibshirani2020pliable}, addresses this challenge by modeling interactions in a structured way. 
It enforces sparsity in both main effects and interactions, while imposing a strong heredity constraint: an interaction can be included only if its corresponding main effect is present. 
This hierarchical structure yields interpretable, stable models and guards against spurious interaction terms, a common risk in high-dimensional settings. 
In its original form, the pliable lasso is estimated via a penalized least squares criterion that combines an $\ell_1$ penalty on main effects with a group lasso ($\ell_2$) penalty on interaction terms. 
Extensions of pliable lasso to different problems beyond continuous response have been considered as in \cite{du2018pliable,asenso2022pliable,asenso2024pliable}.
While effective, this frequentist formulation does not provide a natural mechanism for uncertainty quantification or for incorporating prior knowledge.

In this paper, we propose a Bayesian pliable lasso that extends this framework by placing sparsity-inducing global–local shrinkage priors—such as the horseshoe—on both main and interaction effects. 
These priors strongly shrink irrelevant coefficients toward zero while allowing important effects to escape shrinkage, enabling adaptive regularization in high-dimensional problems. 
The hierarchical prior structure naturally enforces the heredity constraint, while the Bayesian formulation supports principled posterior inference, uncertainty quantification. 
We adapt the pliable model to the Generalized Linear Model (GLM) framework so that it can handle different kinds of responses, such as binary and count data.

Formally, the pliable lasso models the response as
$$
y_i = \beta_0 + \sum_{j=1}^p x_{ij} \left( \beta_j + z_i^\top \bm{\theta}_j \right) + \epsilon_i,
$$
where $x_{ij}$ are main predictors, $z_i$ are modifying variables (often a subset of the predictors), $\beta_j$ are global main effects, and $\bm{\theta}_j \in \mathbb{R}^q$ parameterize how predictor $j$’s effect varies with the modifiers. The strong heredity constraint is expressed as
\begin{equation}
    \label{eq_Plasso_requirement}
    \bm{\theta}_j \textit{ can be nonzero only if } \beta_j \textit{ is nonzero} ,
\end{equation}
ensuring that interactions are present only when their corresponding main effects are included \citep{tibshirani2020pliable}.
In this work, we leverage the Horseshoe prior \citep{carvalho2009handling} to impose sparsity shringkage on both $\beta$ and $ \bm{\theta} $ jointly through the hierarchial structure of the Horseshoe prior.
Horseshoe prior has been widely recognized as default-choice prior for efficiently imposing sparsity \citep{bhadra2019lasso}. It has been adopted in various works, such as \citep{sun2024bhaft,andrinopoulou2016bayesian,mai2024concentration,mai2025hightobit,mai2025handling}.

Our Bayesian formulation retains the pliable lasso’s hierarchical structure but replaces deterministic penalties with priors, enabling a fully probabilistic approach. We develop an efficient Gibbs sampling algorithm that exploits conditional conjugacy for all updates, making posterior computation tractable even in high dimensions. The key idea is to use a reparametrization of horseshoe prior given in \cite{makalic_simple_2016}.
Through simulations and real-data applications, we show that the Bayesian pliable lasso recovers sparse, interpretable interaction structures, and provides coherent measures of uncertainty—advantages that are particularly valuable in complex and noisy domains.

Importantly, we also extend our framework to address situations where the response variable contains missing data, a challenge that is particularly critical in real applications. Our approach treats the missing responses as latent variables, which allows us to derive full conditional distributions in the case of Gaussian observations. Consequently, the proposed Gibbs sampler can be readily applied. Simulation studies demonstrate that this strategy performs effectively in practice.

The rest of the paper is given as follow. In Section \ref{sc_model_method}, we present the model and our pliable Horseshoe approach, together with the Gibbs sampler. Section \ref{sc_missing_data} contain our proposed approach for handling missing data. Simulations studies are presented in Section \ref{sc_simulation}. Application to data from neuroimaging and clinical research on dementia and cognitive decline is presented in Section \ref{sc_real_data}. Additional simulations for binary response are given in Appendix. Conclusion and discussion are given in Section \ref{sc_conclustion}.

\section{Model and Method}
\label{sc_model_method}
\subsection{Model}

Let $y = (y_1, \ldots, y_n)^\top$ denote the response vector of interest. Each response $y_i$ is assumed to arise from a distribution in the exponential family with density
\begin{equation}
\label{eq_GLMmodel}
f(y_i \mid \eta_i ) = \exp\left\{ \frac{y_i \eta_i - b(\eta_i)}{a} + c(y_i) \right\},
\end{equation}
where $\eta_i$ is the canonical parameter, $a$ is a known dispersion constant, and $b(\cdot), c(\cdot)$ are family-specific functions that characterize the distribution (for example, Gaussian, binomial, or Poisson). The choice of $b(\cdot)$ determines both the mean–variance relationship and the natural link function. In particular, the mean response satisfies
$$
\mu_i = \mathbb{E}[y_i \mid X_i, Z_i] = b'(\eta_i),
$$
with variance $\operatorname{Var}(y_i \mid X_i, Z_i) = a b''(\eta_i)$, see \cite{mccullagh1989generalized}.

Following the Pliable Lasso framework of \citet{tibshirani2020pliable}, we extend the linear predictor in generalized linear models to incorporate effect modification by shared covariates $Z_i$. Specifically, we write
\begin{equation}
\label{eq_linear_predictor}
\eta_i = \beta_0 + Z_i^\top \theta_0 + \sum_{j=1}^p x_{ij} \left( \beta_j + Z_i^\top \theta_j \right),
\end{equation}
where:
\begin{itemize}
    \item $\beta_0 \in \mathbb{R}$ is the overall intercept term;

    \item $Z_i \in \mathbb{R}^q$ is the vector of modifying covariates, which act globally across all predictors;

    \item $\theta_0 \in \mathbb{R}^q$ represents the direct effect of modifiers on the outcome;

    \item $\beta_j \in \mathbb{R}$ is the main effect of predictor $x_{ij}$;

    \item $\theta_j \in \mathbb{R}^q$ captures the effect modification of predictor $x_{ij}$ by the shared covariates $Z_i$.
\end{itemize}
This parameterization implies that the contribution of each predictor $x_{ij}$ to the linear predictor is no longer fixed but depends on $Z_i$. In other words, the “slope” of predictor $x_{ij}$ is context-specific and adapts to the modifying variables. This representation allows the model to capture structured heterogeneity in effects, while avoiding the combinatorial explosion of including all possible pairwise interactions.

For notational convenience, let $X \in \mathbb{R}^{n \times p}$ denote the design matrix of predictors, and $Z \in \mathbb{R}^{n \times q}$ the matrix of modifying covariates. The mean response is then linked to the predictor via a chosen link function $g: \mathbb{R} \to \mathbb{R}$,
\begin{equation}
\label{eq_link_function}
g\big(\mathbb{E}[y_i \mid X_i, Z_i]\big) = \eta_i.
\end{equation}

\begin{rmk}
This formulation generalizes classical GLMs in two important ways:
\begin{itemize}
    \item When $\theta_0 = \theta_j = 0$ for all $j$, the model reduces to a standard GLM with predictors $X$.

    \item When $\theta_j \neq 0$, the model introduces structured interactions between $X$ and $Z$, where $Z$ serves as a common modifier across all predictors. This provides a parsimonious approach to modeling effect heterogeneity, compared with including all $X \times Z$ interactions explicitly.
\end{itemize}
\end{rmk}

The model accommodates a wide range of response types depending on the choice of exponential family and link. For example: Gaussian responses with identity link, binary responses with logit/probit link, and count responses with log link. The pliable structure makes the model especially suitable for high-dimensional problems where interaction effects are expected but difficult to estimate without imposing structure.

The inclusion of shared modifiers $Z$ achieves a form of dimension reduction in interaction modeling: rather than estimating $pq$ unrestricted interaction terms, the structure in \eqref{eq_Plasso_requirement} and \eqref{eq_linear_predictor} allows these interactions to be expressed through low-dimensional modifier vectors, improving interpretability and statistical efficiency. This is particularly relevant in settings where effect modification is biologically or contextually motivated, such as gene–environment interactions, treatment–covariate interactions, or socio-demographic effect heterogeneity, as discused in \cite{tibshirani2020pliable}.

\subsection{Bayesian pliable Lasso using Horseshoe prior}

Let
$$
\Psi = \{\beta_0,\ \theta_0,\ \boldsymbol\beta=(\beta_1,\dots,\beta_p),\ \Theta=(\theta_1,\dots,\theta_p)
\}
,
$$
and write the likelihood 
$
L(\mathbf y\mid\Psi)=\prod_{i=1}^n f(y_i\mid\eta_i)
$ with $\eta_i$ given by \eqref{eq_linear_predictor}.
Using $\pi_{\mathrm{HS}}$ to denote the joint Horseshoe prior on $(\boldsymbol\beta,\Theta)$ and $\pi(\beta_0,\theta_0)$ for the remaining priors, the posterior distribution in abstract form is given as
$$
\pi(\Psi\mid \mathbf y, X, Z)
\;\propto\;
L(\mathbf y\mid \Psi)\;\pi_{\mathrm{HS}}\bigl(\boldsymbol\beta,\Theta\bigr)\;
\pi(\beta_0,\theta_0).
$$

We use Horseshoe prior \citep{carvalho2009handling}
to enforce sparsity on both $\beta_j$ and $\theta_j$, using shared local and global scale parameters for hierarchical shrinkage. Let IG denote the inverse-gamma distribution. The Horseshoe prior in our problem is given as
\begin{equation}
    \label{eq_HS_prior}
    \begin{aligned}
\beta_j & \sim \mathcal{N}(0, \lambda_j^2   \tau^2) ; 
\\
\bm{\theta}_j & \sim \mathcal{N}(0, \lambda_j^2  \tau^2 I_q) 
\\
\lambda_j &\sim \text{Cau}^+(0, 1)
,\,\,\,
 \tau  \sim \text{Cau}^+(0, 1) 
\end{aligned}
\end{equation}
for other parameter
$$
\bm{\theta}_0 \sim \mathcal{N}(0, \sigma_0^2 I_q)
, \quad 
\beta_0 \sim \mathcal{N}(0, \sigma_0^2)
$$
for $j = 1, \dots, p$, where $\mathrm{Cau}^+(0, 1)$ denotes the standard half-Cauchy distribution, 
truncated to the positive real line, with density proportional to $(1 + v^2)^{-1} \mathbf{1}_{(0, \infty)}(v)$. In the case of Gaussian noise, we put the prior on the noise variance as $\sigma^2 \sim \text{IG}(a_0, b_0) $. We fix $ \sigma_0^2 =1, a_0 = b_0 = 10^{-2} $ in our algorithms.

The Horseshoe prior is a powerful method for sparse Bayesian modeling with high-dimensional data, utilizing a hierarchical global-local shrinkage structure. It employs a single global parameter, $ \tau $, to control the overall level of shrinkage, while local parameters, $ \lambda_j $, allow for coefficient-specific adaptation. This design is highly effective because it aggressively shrinks noise by pulling small or irrelevant coefficients towards zero, yet simultaneously protects large, important signals from being overly shrunk due to the heavy-tailed nature of its half-Cauchy distribution. As a result, the Horseshoe prior behaves much like a continuous spike-and-slab prior, offering both computational efficiency and theoretically optimal performance in sparse regimes \citep{van2017adaptive}..

In the pliable Horseshoe setting, we extend this global–local framework beyond the regression coefficients $ \beta$  to also govern the modifier effects $ \bm{\theta}_j$, this is significantly different to \cite{sun2024bhaft} which simply employs Horseshoe prior separately. 
By sharing the same hierarchical structure across both $ \beta$  and $ \bm{\theta}_j$ , the prior automatically enforces the coupling required by the pliable lasso constraint as given in \eqref{eq_Plasso_requirement}. 
Intuitively, when a covariate effect $ \beta_j$  is shrunk toward zero, its corresponding modifier effects are simultaneously shrunk, while strong signals are preserved across both levels. This joint shrinkage mechanism makes the pliable Horseshoe a natural Bayesian analogue to the pliable lasso.

\subsection{Gibbs sampler}
\label{sc_gibbssampler}
To enable Gibbs sampling, we reparameterize the half-Cauchy priors using auxiliary variables \citep{makalic_simple_2016}:
$$
\lambda_j^2 \sim \text{IG}(1/2, 1/\nu_j), \quad \nu_j \sim \text{IG}(1/2, 1) 
.
$$
$$
\tau^2 \sim \text{IG}(1/2, 1/\xi), \quad \xi \sim \text{IG}(1/2, 1)
.
$$
For linear model,
$$
y = \beta_0 \mathbf{1}_n + Z \theta_0 + \sum_{j=1}^p x_j (\beta_j + Z \theta_j) + \varepsilon, \quad \varepsilon \sim \mathcal{N}(0, \sigma^2 I_n).
$$

 Gibbs Sampling Steps for the Linear Pliable Lasso with Horseshoe Prior: 
At each iteration of the Gibbs sampler, the following updates are performed:

1. Update regression coefficients $\beta_j$.

For each predictor $j=1,\dots,p$, define the partial residual excluding predictor $j$:
$$
r^{(-j)} = y - \beta_0 - Z \theta_0 - \sum_{k \ne j} x_k (\beta_k + Z \theta_k),
$$
and let $w_j = x_j$, $Z_j = \mathrm{diag}(x_j) Z$.
The full conditional distribution of $\beta_j$ is Gaussian:
$$
\beta_j \mid \cdot \;\sim\; \mathcal{N}\!\left(\mu_{\beta_j}, V_{\beta_j}\right),
$$
with
$$
V_{\beta_j} = \left( \frac{w_j^\top w_j}{\sigma^2} + \frac{1}{\lambda_j^2 \tau^2} \right)^{-1}, 
\quad 
\mu_{\beta_j} = V_{\beta_j} \cdot \frac{w_j^\top \big(r^{(-j)} - Z_j \theta_j\big)}{\sigma^2}.
$$

2. Update modifier effects $\theta_j$.

Conditional on $\beta_j$, define:
$$
r^{(-j)} = y - \beta_0 - Z \theta_0 - \sum_{k \ne j} x_k (\beta_k + Z \theta_k) - x_j \beta_j,
\quad 
Z_j = \mathrm{diag}(x_j) Z.
$$
The full conditional for $\theta_j \in \mathbb{R}^q$ is Gaussian:
$$
\theta_j \mid \cdot \;\sim\; \mathcal{N}\!\left(\mu_{\theta_j}, V_{\theta_j}\right),
$$
with
$$
V_{\theta_j} = \left( \frac{Z_j^\top Z_j}{\sigma^2} + \frac{I_q}{\lambda_j^2 \tau^2} \right)^{-1}, 
\quad 
\mu_{\theta_j} = V_{\theta_j} \cdot \frac{Z_j^\top r^{(-j)}}{\sigma^2}.
$$

3. Update local shrinkage parameters $\lambda_j^2, \nu_j$.

Through the scale mixture representation, the full conditionals for local shrinkage parameters are inverse-gamma:
$$
\lambda_j^2 \mid \cdot \;\sim\; \mathrm{IG}\!\left( \frac{q+1+1}{2}, \; \frac{1}{\nu_j} + \frac{\beta_j^2 + \|\theta_j\|_2^2}{2 \tau^2} \right),
\quad
\nu_j \mid \cdot \;\sim\; \mathrm{IG}\!\left(\tfrac{1}{2}, \; 1 + \frac{1}{\lambda_j^2}\right).
$$

4. Update the global scale $\tau^2$ and auxiliary parameter $\xi$.

The horseshoe prior also introduces a global shrinkage parameter. Its full conditionals are:
$$
\tau^2 \mid \cdot \;\sim\; \mathrm{IG}\!\left( \frac{p(q+1) + 1}{2}, \; \frac{1}{\xi} + \frac{1}{2} \sum_{j=1}^p \frac{\beta_j^2 + \|\theta_j\|_2^2}{\lambda_j^2} \right),
\quad
\xi \mid \cdot \;\sim\; \mathrm{IG}\!\left(\tfrac{1}{2}, \; 1 + \frac{1}{\tau^2}\right).
$$

5. Update intercept terms $(\beta_0, \theta_0)$.

The intercept block consists of the main intercept $\beta_0$ and the modifier intercepts $\theta_0 \in \mathbb{R}^q$. Both have Gaussian priors and therefore admit Gaussian full conditionals.

* For $\beta_0$, define
$
r_{\beta_0} = y - \sum_{j=1}^p x_j(\beta_j + Z \theta_j) - Z \theta_0,
$
and sample
$$
\beta_0 \mid \cdot \;\sim\; \mathcal{N}(\mu_0, V_0),
\quad 
V_0 = \left(\frac{n}{\sigma^2} + \frac{1}{\sigma_0^2} \right)^{-1}, 
\quad 
\mu_0 = V_0 \cdot \frac{\sum_{i=1}^n r_{\beta_0,i}}{\sigma^2}.
$$

* For $\theta_0$, define
$
r_{\theta_0} = y - \beta_0 - \sum_{j=1}^p x_j(\beta_j + Z \theta_j),
$
and sample
$$
\theta_0 \mid \cdot \;\sim\; \mathcal{N}(\mu_{\theta_0}, V_{\theta_0}), 
\quad 
V_{\theta_0} = \left( \frac{Z^\top Z}{\sigma^2} + \frac{I_q}{\sigma_0^2} \right)^{-1}, 
\quad 
\mu_{\theta_0} = V_{\theta_0} \cdot \frac{Z^\top r_{\theta_0}}{\sigma^2}.
$$

6. Update noise variance $\sigma^2$.
Finally, the residual variance is updated using its inverse-gamma full conditional. With residuals $r = y - \mu$,
$$
\sigma^2 \mid \cdot \;\sim\; \mathrm{IG}\!\left(a_0 + \frac{n}{2}, \; b_0 + \frac{1}{2}\|r\|_2^2 \right).
$$

This scheme cycles through all parameter blocks, yielding posterior draws of regression coefficients, modifier effects, and shrinkage parameters. The  horseshoe prior adaptively shrinks irrelevant predictors while allowing important predictors and interactions to remain large.

The proposed method has been implemented in the \texttt{R} package \texttt{hspliable}, available on GitHub (\url{https://github.com/tienmt/hspliable}
). The implementation leverages \texttt{Rcpp} and \texttt{RcppArmadillo} to achieve computational efficiency.

\section{Handling missing data in the response}
\label{sc_missing_data}

When some outcomes in the response vector $y=(y_1,\dots,y_n)$ are missing; we treat the missing entries as latent variables and perform data augmentation inside the Gibbs sampler. Let $\mathcal{O}=\{i: y_i\ \text{is observed}\}$ and $\mathcal{M}=\{i: y_i\ \text{is missing}\}$. The complete-data model is
$$
y_i = \beta_0 + z_i^\top\theta_0 + \sum_{j=1}^p x_{ij}\big(\beta_j + z_i^\top\theta_j\big) + \varepsilon_i,
\quad
\varepsilon_i\sim\mathcal{N}(0,\sigma^2),
$$
for $i=1,\dots,n$. Under this model, the missing responses $y_{\mathcal{M}}$ are additional unknowns; the sampler targets the joint posterior
$$
p(\text{parameters},\,y_{\mathcal{M}}\mid y_{\mathcal{O}},X,Z).
$$
The conditional distribution used to impute a missing response is simply the Gaussian likelihood evaluated at that index, conditional on the current parameter values. 

Concretely, given current draws $\beta_0^{(t-1)},\theta_0^{(t-1)},\{\beta_j^{(t-1)},\theta_j^{(t-1)}\}_{j=1}^p$ and $\sigma^{2\,(t-1)}$, the conditional mean for an index $i\in\mathcal{M}$ is
$$
\mu_i^{(t-1)} \;=\; \beta_0^{(t-1)} + z_i^\top\theta_0^{(t-1)}
+ \sum_{j=1}^p x_{ij}\big(\beta_j^{(t-1)} + z_i^\top\theta_j^{(t-1)}\big),
$$
and the conditional variance is $\sigma^{2\,(t-1)}$. The reason is immediate: the likelihood factor for $y_i$ is $\mathcal{N}(\mu_i,\sigma^2)$, and because $y_i$ does not appear elsewhere except through that Gaussian likelihood, the full conditional for $y_i$ given the parameters is
$$
p(y_i\mid\text{parameters},y_{\mathcal{O}})\;=\;\mathcal{N}\big(\mu_i,\sigma^2\big).
$$
Thus the imputation step at iteration $t$ draws
$$
y_i^{(t)}\sim\mathcal{N}\big(\mu_i^{(t-1)},\,\sigma^{2\,(t-1)}\big),\qquad i\in\mathcal{M}.
$$
After imputing all missing entries, we form the completed response vector $\tilde y^{(t)}$ with $\tilde y^{(t)}_{\mathcal{O}}=y_{\mathcal{O}}$ and $\tilde y^{(t)}_{\mathcal{M}}=y^{(t)}_{\mathcal{M}}$. The remaining Gibbs updates, from Section~\ref{sc_gibbssampler}, for the model parameters proceed unchanged by replacing the original $y$ with $\tilde y^{(t)}$. Because the conditional distributions in the sampler are derived from the Gaussian complete-data likelihood, no algebraic modification of those conditionals is necessary: each conditional has the same form as in the fully observed case, evaluated at the completed data.

We remark that this data-augmentation approach implicitly assumes that the missingness mechanism is ignorable (for example, missing at random conditional on observed covariates and model parameters). If missingness depends on unobserved quantities not modelled here, then the imputation model would need to be enlarged to include an explicit model for the missingness mechanism.

Overall, the only change required to the Gibbs sampler described earlier is the addition of an imputation step at the start of each iteration. All parameter conditional distributions remain the same; the imputed $y_{\mathcal{M}}^{(t)}$ are treated as additional latent draws and may be stored if one wishes to report posterior imputed values or to perform posterior predictive checks.

\section{Simulation studies}
\label{sc_simulation}
\subsection{Setup}

We design simulation studies under a variety of configurations for the predictor matrix $X$ and the modifying covariates $Z$, allowing both binary and continuous cases.

\begin{table}[!h]
	\centering
	\caption{Outline of simulation settings.}
	\begin{tabular}{ | l l l l  | }
\hline
Setting & Name &  $Z_{ij} $ & $X_i $ 
		\\
		\hline
I & both continuous 	& $ \mathcal{N}(0, 1) $ & $\mathcal{N}(0, \mathbb{I}_p ) $  
		\\
II & binary, continuous 	& $ {\rm Ber}(0.5) $ & $\mathcal{N}(0, \mathbb{I}_p ) $  
		\\
III & continuous, correlated & $ \mathcal{N}(0, 1) $ & $\mathcal{N}(0, \Sigma) $  
		\\
IV & binary, correlated & $ {\rm Ber}(0.5) $ & $\mathcal{N}(0, \Sigma) $  
		\\
V & continuous, binary & $ \mathcal{N}(0, 1) $  & $ {\rm Ber}(0.5) $  
		\\
VI & both binary & $ {\rm Ber}(0.5) $ & $ {\rm Ber}(0.5) $  
		\\
		\hline
	\end{tabular}
	\label{tb_simu_settings}
\end{table}

For continuous predictors, we generate each row $X_i \sim \mathcal{N}(0, \Sigma)$, where $\Sigma$ governs the correlation structure among predictors. Two covariance structures are considered: (i) an independence structure with $\Sigma = \mathbb{I}_p$, and (ii) a correlated structure with autoregressive form $(\Sigma)_{ij} = \rho_X^{|i-j|}$ for all $i, j$. We set the number of predictors to $p=10$ and the number of modifying covariates to $q=4$. The baseline parameters are fixed at $\beta_0 = 1$ and $\theta_0 = (-0.5, -0.5, -0.5, -0.5)^\top$.
The true regression coefficients are specified as
$
\beta = (2,\,-2,\,2,\,2,\,0,\,\ldots,\,0)^\top,
$
and the true effect modification matrix $\Theta = [\theta_1, \ldots, \theta_p]$ is defined such that
$$
\theta_1 = (1,\,1,\,1,\,1)^\top,\quad 
\theta_2 = (-2,\,-2,\,-2,\,-2)^\top,\quad 
\theta_3 = (1,\,2,\,3,\,4)^\top,
$$
with all remaining $\theta_j$ set to zero.
We vary the sample size across $n \in \{200, 500, 1000\}$. 
The modifying covariates $Z$ are generated either from the standard normal distribution, $\mathcal{N}(0,1)$, or from a Bernoulli distribution with success probability 0.5. 
Finally, the noise term is drawn independently from $\mathcal{N}(0,1)$.
A summary of the simulation settings is provided in Table~\ref{tb_simu_settings}.

We compare our pliable Horseshoe method (denoted pHS) with pliable lasso (denoted pLasso) available from the \texttt{R} package \texttt{svreg} on Github (\url{https://github.com/Tanya-Garcia-Lab/svreg/}); Lasso from the \texttt{R} package \texttt{glmnet}, \citep{glmnet}; and the simple Horseshoe method available in the \texttt{R} package \texttt{horseshoe} \citep{horseshoe_package}.
The pliable lasso and lasso are run with 3-fold cross validation to select the best tuning parameter. Our proposed pliable Horseshoe is run with 5000 steps and the first 500 steps are removed as burnin. Horseshoe is run with with 5000 steps and the first 1000 steps are removed as burnin.

\begin{figure}[!ht]
    \centering
    \includegraphics[width=16cm]{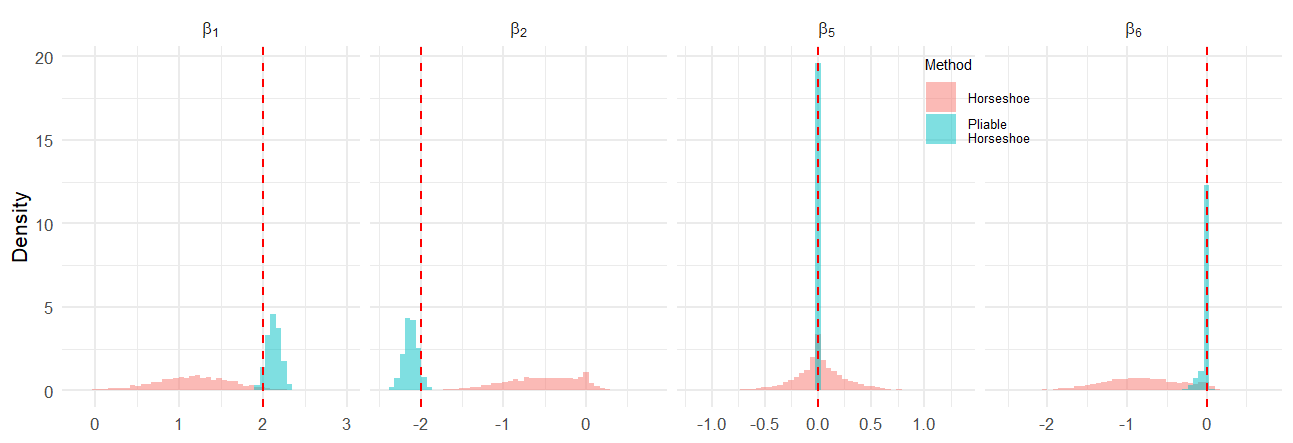}
    \caption{\it Histogram plots for comparison of posterior distributions by parameter for Hoseshoe and pliable Horseshoe methods in Setting I with $ n=200,p=10,q=4$. The true value for $\beta_1$ is $2$, for $\beta_2$ is $-2$, and for $\beta_5 , \beta_6 $ are $0$.}
    \label{fig:posterior_example1256}
\end{figure}

We evaluate the considered methods using estimation error as 
$$ 
{\rm Est.} (\beta) := \| \widehat{\beta} - \beta \|_2^2, 
\quad
{\rm Est.} (\bm{\theta}) := \| \widehat{\bm\theta} - \bm{\theta} \|_2^2
,
$$
where $ \widehat{\beta}$ and $\widehat{\theta} $ are the estimate from the considered methods. We also access the prediction performance of the considered methods using prediction error on testing data,
$$
{\rm Pred} := 
\frac{1}{n_{\rm test}} \sum_{i =1}^{n_{\rm test}} (y_{{\rm test},i} - \widehat{y}_i)^2
$$
where $\widehat{y}_i = \widehat{\beta}_0 + Z_{{\rm test}, i}^\top \widehat{\theta}_0 + \sum_{j=1}^p X_{{\rm test},i} \left( \widehat{\beta}_j + Z_{{\rm test},i}^\top \widehat{\theta}_j \right) $ is the prediction on testing data for pliable lasso and pliable horseshoe.
For lasso and horseshoe method the prediction is without the the interaction part. There, $ y_{{\rm test}}, Z_{{\rm test}} $ and $ X_{{\rm test}} $ are testing dataset simulated as the training data for each simulation and we fix $ n_{\rm test} = 50 $ in all settings.
In addition, we also evaluate the variable selection performance of the considered methods. 
Specifically, we compute the standard variable selection metrics based on the counts of true positives (TP), false negatives (FN), false positives (FP), and true negatives (TN). The following measures: accuracy, false dicovery rate, false positive rate, are considered:
$$
\displaystyle \text{Accuracy} = \frac{\text{TP + TN}}{\text{TP + FP + FN + TN}}
; \quad
\displaystyle \text{FDR} = \frac{\text{FP}}{\text{TP + FP} }
; \quad
\displaystyle \text{FPR} = \frac{\text{FP}}{ \text{FP + TN}}
.
$$
Each simulation setting are repeated 100 times and we report the average result together with its standard deviation. The results are given in Table \ref{tb_seting_1_2}, \ref{tb_seting_3_4} and \ref{tb_seting_5_6}.

We note that the original study on the pliable lasso \citep{tibshirani2020pliable} did not provide results on estimation error or variable selection sensitivity, as their attention was limited to prediction error.

\begin{table}[!ht]
\centering
	\caption{Simulation results for $ p = 10,q = 4 $ for Setting I and II. HS: Horseshoe; pLasso: Pliable lasso; pHS: pliable Horseshoe. Note that for HS and Lasso we do not have estimation for $\bm{\theta} $.}
	\begin{tabular}{ | r | ccc  ccc | }
		\hline \hline
Method  & ${\rm Est.} (\beta) $
& ${\rm Est.} (\bm{\theta}) $
& {\rm Pred} 
&  Accuracy
& FDR
& FPR
		\\
		\hline
 \multicolumn{7}{| c | }{ Setting I, $n = 200$ }
\\ 
\hline
HS 
& 2.95 (2.25) & \_ & 53.6 (19.1) 
& 0.92 (0.10) & 0.02 (0.06) & 0.01 (0.04)
		\\
Lasso
& 3.03 (2.61) & \_ & 53.7 (19.0) 
& 0.69 (0.17) & 0.39 (0.18) & 0.51 (0.30)
\\
pLasso
& 0.32 (0.33) & 0.48 (0.24) & 1.62 (0.40)
& 0.44 (0.06) & 0.58 (0.03) & 0.93 (0.10)
		\\
pHS
& 0.05 (0.02) & 0.22 (0.07) & 1.24 (0.23) 
& 0.98 (0.05) & 0.04 (0.09) & 0.04 (0.08)
		\\
		\hline
 \multicolumn{7}{ | c | }{  Setting I,  $n = 500$ }
\\ 
\hline
HS 
& 0.97 (0.52) &  & 51.2 (17.9) 
& 0.99 (0.03) & 0.01 (0.04) & 0.01 (0.03)
		\\
Lasso
& 1.06 (0.63) &  & 51.3 (17.8) 
& 0.68 (0.16) & 0.42 (0.14) & 0.54 (0.27)
		\\
pLasso
& 0.13 (0.12) & 0.16 (0.10) & 1.32 (0.27) 
& 0.57 (0.12) & 0.51 (0.08) & 0.72 (0.20)
		\\
pHS
& 0.02 (0.01) & 0.07 (0.01) & 1.10 (0.25) 
& 0.98 (0.05) & 0.04 (0.09) & 0.03 (0.08)
		\\
		\hline
 \multicolumn{7}{ | c | }{  Setting I,  $n = 1000$ }
\\ 
\hline
HS 
& 0.48 (0.31) &  & 54.6 (22.3) 
& 1.00 (0.02) & 0.01 (0.04) & 0.01 (0.04)
		\\
Lasso
& 0.56 (0.41) &  & 54.8 (22.5) 
& 0.69 (0.17) &  0.40 (0.16) & 0.51 (0.28) 
		\\
pLasso
& 0.06 (0.06) & 0.10 (0.07) & 0.93 (0.07)
& 0.74 (0.11) & 0.37 (0.11) & 0.43 (0.19)
		\\
pHS
& 0.01 (0.00) & 0.03 (0.01) & 1.06 (0.20)
& 0.99 (0.03) & 0.02 (0.06) & 0.02 (0.05)
		\\
\hline		\hline
 \multicolumn{7}{ | c | }{  Setting II,  $n = 200$ }
\\ 
\hline
HS 
& 44.9 (5.86) & & 15.2  (5.04)
& 0.99 (0.03) & 0.02 (0.06) & 0.02 (0.05)
		\\
Lasso
& 41.9  (5.74) & & 15.3 (5.04)
& 0.68 (0.16) & 0.42 (0.14) & 0.54 (0.27)
		\\
pLasso
& 43.6 (4.77) & 14.1  (1.00) & 15.2 (5.46)
& 0.44 (0.06) & 0.58 (0.03) & 0.93 (0.10)
		\\
pHS
& 0.21 (0.11) & 0.80 (0.27) &  1.27 (0.27)
& 0.98 (0.04) & 0.03 (0.07) & 0.03 (0.06)
		\\
		\hline
 \multicolumn{7}{ | c| }{ Setting II, $n = 500$ }
\\ 
\hline
HS 
& 44.5 (3.35) & & 13.8  (4.63)
& 0.99 (0.03) & 0.02 (0.07) & 0.02 (0.06)
		\\
Lasso
& 42.5 (3.26) & & 13.8  (4.67)
& 0.70 (0.17) & 0.40 (0.15) & 0.51 (0.28)
		\\
pLasso
& 44.2 (3.35) & 13.5 (0.64) & 15.1 (4.51)
& 0.55 (0.10) & 0.52 (0.06) & 0.74 (0.17)
		\\
pHS
& 0.07 (0.04) & 0.25 (0.07) & 1.06 (0.23)
& 0.99 (0.03) & 0.01 (0.05) & 0.01 (0.04)
		\\
		\hline
 \multicolumn{7}{ |c| }{ Setting II, $n = 1000$ }
\\ 
\hline
HS 
& 44.9  (2.48) & & 13.4  (4.81)
& 0.99 (0.03) & 0.01 (0.05) & 0.01 (0.04)
		\\
Lasso
& 43.4 (2.56) & & 13.4  (4.81)
& 0.69 (0.17) & 0.40 (0.16) & 0.52 (0.29)
		\\
pLasso
& 44.3 (2.45) & 13.5 (0.46) & 14.6 (3.65)
& 0.72 (0.12) & 0.39 (0.11) & 0.46 (0.20)
		\\
pHS
& 0.03 (0.02) & 0.12 (0.03) & 1.04 (0.21)
& 0.99 (0.02) & 0.01 (0.05) & 0.01 (0.04)
		\\
\hline	\hline	
\end{tabular}
\label{tb_seting_1_2}
\end{table}

\subsection{Simulation results}

A general result from our simulations is that the pliable Horseshoe outperform the pliable lasso in all consider settings in estimation, prediction and variable selection.

Additional simulations for binary responses with logistic regression are given Appendix.

\paragraph*{Result in estimation accuracy}

As shown in Tables \ref{tb_seting_1_2}, \ref{tb_seting_3_4}, and \ref{tb_seting_5_6}, the pliable Horseshoe consistently yields the smallest estimation error when compared with the other three competing methods. 
This advantage is especially notable in Settings II, IV, and VI, where the modifying covariate $Z$ is binary. In these cases, the pliable Horseshoe is the only method that continues to provide stable and accurate estimates, whereas the other approaches fail to adequately capture the interaction structure. 
In contrast, for Settings I, III, and V, where $Z$ is continuous, the pliable lasso performs reasonably well, but still does not reach the same level of accuracy as the pliable Horseshoe.
Nonetheless, the pliable lasso achieves considerably better results than either the naive lasso or the naive Horseshoe (which do not account for interactions), confirming that explicitly modeling interactions leads to substantial gains. 
These findings underscore the importance of adopting methods that are specifically designed to handle effect modification, a point that has also been emphasized in \cite{tibshirani2020pliable}.

\begin{table}[!ht]
\centering
	\caption{Simulation results for $ p = 10,q = 4 $ for Setting III and IV. HS: Horseshoe; pLasso: Pliable lasso; pHS: pliable Horseshoe. Note that for HS and Lasso we do not have estimation for $\bm{\theta} $.}
	\begin{tabular}{ | l | ccc  ccc | }
		\hline \hline
Method  & ${\rm Est.} (\beta) $
& ${\rm Est.} (\bm{\theta}) $
& {\rm Pred} 
&  Accuracy
& FDR
& FPR
		\\
		\hline
 \multicolumn{7}{ |c| }{ Setting III, $n = 200$ }
\\ 
\hline
HS 
& 2.71 (2.33) & \_ &  31.2 (11.9)
& 0.92 (0.10) & 0.01 (0.05) & 0.00 (0.03)
		\\
Lasso
& 3.20 (2.81) & \_ &  31.5 (11.7)
& 0.68 (0.17) & 0.39 (0.18) & 0.51 (0.31)
		\\
pLasso
&  0.39 (0.36) & 0.68 (0.37) & 1.85 (0.49) 
&  0.47 (0.08) & 0.57 (0.04) & 0.89 (0.14) 
		\\
pHS
& 0.08 (0.05) & 0.32 (0.09) & 1.30 (0.28) 
& 1.00 (0.01)  & 0.00 (0.02) & 0.00 (0.02)
		\\
		\hline
 \multicolumn{7}{ |c }{ Setting III,  $n = 500$ }
\\ 
\hline
HS 
& 0.83 (0.63) & \_ &  31.4 (12.4)
& 0.99 (0.03)  & 0.01 (0.05) & 0.01 (0.05)
		\\
Lasso
& 0.98 (0.80) & \_ & 31.6 (12.4)
& 0.63 (0.17)  & 0.45 (0.14) & 0.62 (0.28)
		\\
pLasso
& 0.14 (0.16) & 0.37 (0.23) & 1.38 (0.38)
&  0.59 (0.11) & 0.49 (0.07) & 0.68 (0.18)
		\\
pHS
& 0.03 (0.01) & 0.11 (0.03) & 1.09 (0.22)
&  0.99 (0.04) & 0.03 (0.08) & 0.02 (0.07)
		\\
		\hline
 \multicolumn{7}{ |c | }{ Setting III,  $n = 1000$ }
\\ 
\hline
HS 
& 0.46 (0.29) & \_ &  28.9  (9.25)
&  1.00 (0.01) & 0.00 (0.03) & 0.00 (0.02)
		\\
Lasso
& 0.59 (0.40) & \_ &  29.0  (9.27)
& 0.66 (0.17)  & 0.43 (0.14) & 0.57 (0.28)
		\\
pLasso
& 0.08 (0.10) & 0.25 (0.12) & 1.35 (0.27)
& 0.72 (0.12)  & 0.39 (0.12) & 0.46 (0.21)
		\\
pHS
& 0.01 (0.01) & 0.05 (0.01) &  1.06 (0.23)
& 0.99 (0.04) & 0.03 (0.08) & 0.02 (0.07)
		\\
\hline		\hline
 \multicolumn{7}{ |c | }{ Setting IV, $n = 200$ }
\\ 
\hline
HS 
& 44.6 (5.99) & \_ & 8.69 (2.55)
& 0.99 (0.04) & 0.02 (0.07) & 0.02 (0.06)
		\\
Lasso
& 40.8 (6.54) & \_ & 8.78 (2.57)
& 0.64 (0.16) & 0.45 (0.13) & 0.60 (0.27)
		\\
pLasso
& 43.1 (6.08) & 15.6  (1.39) & 8.16 (3.18)
& 0.46 (0.08) & 0.57 (0.04) & 0.90 (0.13)
		\\
pHS
& 0.33 (0.22) & 1.11 (0.37) & 1.25 (0.26)
& 0.99 (0.04) & 0.03 (0.07) & 0.02 (0.07)
		\\
		\hline
 \multicolumn{7}{ |c| }{ Setting IV, $n = 500$ }
\\ 
\hline
HS 
& 45.0 (3.99) & \_ & 8.28 (2.22)
& 0.99 (0.04) & 0.02 (0.07) & 0.02 (0.06)
		\\
Lasso
& 42.5 (4.11) & \_ & 8.30 (2.20)
& 0.64 (0.16) & 0.45 (0.13) & 0.60 (0.27)
		\\
pLasso
& 44.3 (3.42) & 15.0  (0.81) & 8.35 (2.57)
& 0.60 (0.10) & 0.49 (0.07) & 0.68 (0.17)
		\\
pHS
& 0.11 (0.06) & 0.39 (0.12) & 1.11 (0.26)
& 0.99 (0.04) & 0.02 (0.07) & 0.02 (0.07)
		\\
		\hline
 \multicolumn{7}{ |c| }{ Setting IV,  $n = 1000$ }
\\ 
\hline
HS 
& 44.7  (2.90) & \_ & 7.97 (2.33)
& 0.99 (0.03) & 0.01 (0.05) & 0.01 (0.05)
		\\
Lasso
& 42.9 (2.88) & \_ & 7.99 (2.34)
& 0.67 (0.16) & 0.42 (0.14) & 0.55 (0.26)
		\\
pLasso
& 44.1 (2.53) & 14.9  (0.55) & 7.34 (1.66)
& 0.74 (0.11) & 0.37 (0.11) & 0.43 (0.18)
		\\
pHS
& 0.05 (0.04) & 0.18 (0.06) & 1.04 (0.20)
& 0.99 (0.03) & 0.02 (0.06) & 0.01 (0.05)
		\\
\hline	\hline	
\end{tabular}
\label{tb_seting_3_4}
\end{table}

\paragraph*{Result in prediction error}

Similarly, as shown in Tables \ref{tb_seting_1_2}, \ref{tb_seting_3_4}, and \ref{tb_seting_5_6}, the pliable Horseshoe consistently achieves the lowest prediction error among the four competing methods. 
In addition, the prediction error decreases as the sample size increases, demonstrating the method’s ability to effectively utilize additional data. A particularly striking result arises in Settings II, IV, and VI, where the modifying covariate 
$Z$ is binary. 
In these cases, all other approaches—including the pliable lasso—fail to capture the underlying structure and result in poor prediction performance, whereas our proposed pliable Horseshoe continues to maintain a small prediction error. 
This highlights the robustness of the Bayesian formulation in challenging scenarios where traditional approaches may break down.

\begin{table}[!ht]
\centering
	\caption{Simulation results for $ p = 10,q = 4 $ in Setting V and VI. HS: Horseshoe; pLasso: Pliable lasso; pHS: pliable Horseshoe. Note that for HS and Lasso we do not have estimation for $\bm{\theta} $.}
	\begin{tabular}{ | l | ccc  ccc | }
		\hline \hline
Method  & ${\rm Est.} (\beta) $
& ${\rm Est.} (\bm{\theta}) $
& {\rm Pred} 
&  Accuracy
& FDR
& FPR
		\\
		\hline
 \multicolumn{7}{ |c| }{ Setting V, $n = 200$ }
\\ 
\hline
HS 
& 2.35 (1.59) & \_ &  17.0  (4.65)
& 0.94 (0.08) & 0.01 (0.03) & 0.00 (0.03)
		\\
Lasso
& 2.53 (1.70) & \_ & 18.6  (5.03) 
& 0.69 (0.18) & 0.39 (0.17) & 0.51 (0.30)
		\\
pLasso
& 0.46 (0.37) & 1.02 (0.43) &  1.42 (0.33) 
& 0.45 (0.07) & 0.58 (0.03) & 0.91 (0.12)
		\\
pHS
& 0.20 (0.11) & 0.86 (0.20) & 1.40 (0.34) 
& 0.98 (0.04) & 0.03 (0.07) & 0.02 (0.06)
		\\
		\hline
 \multicolumn{7}{ |c| }{ Setting V, $n = 500$ }
\\ 
\hline
HS 
& 0.95 (0.51) & \_ &  15.8  (5.18)
& 0.99 (0.03) & 0.02 (0.06) & 0.01 (0.05)
		\\
Lasso
& 1.10 (0.69) & \_ &  16.7  (5.50)
& 0.69 (0.18) & 0.40 (0.17) & 0.52 (0.30)
		\\
pLasso
& 0.17 (0.13) & 0.60 (0.18) &  1.20 (0.23)
& 0.56 (0.11) & 0.52 (0.07) & 0.74 (0.19)
		\\
pHS
& 0.07 (0.04) & 0.28 (0.08) &  1.11 (0.23)
& 0.98 (0.04) & 0.03 (0.07) & 0.02 (0.06)
		\\
		\hline
 \multicolumn{7}{ |c| }{ Setting V, $n = 1000$ }
\\ 
\hline
HS 
& 0.44 (0.27) & \_ & 15.6 (4.37) 
& 0.99 (0.03) & 0.02 (0.06) & 0.01 (0.05)
		\\
Lasso
& 0.50 (0.30) & \_ &  16.2  (4.42)
& 0.68 (0.18) & 0.40 (0.17) & 0.54 (0.30)
		\\
pLasso
& 0.07 (0.07) & 0.51 (0.12) & 1.14 (0.21) 
& 0.71 (0.11) & 0.40 (0.10) & 0.48 (0.19)
		\\
pHS
& 0.03 (0.02) & 0.13 (0.03) &  1.05 (0.21)
& 0.98 (0.04) & 0.03 (0.07) & 0.03 (0.06)
		\\
\hline		\hline
 \multicolumn{7}{ |c | }{ Setting VI, $n = 200$ }
\\ 
\hline
HS 
& 45.0  (5.19) & \_ & 4.77 (1.01)
& 1.00 (0.01) & 0.00 (0.03) & 0.00 (0.02)
		\\
Lasso
& 41.3  (5.10) & \_ & 4.92 (1.05)
& 0.70 (0.17) & 0.39 (0.16) & 0.51 (0.29)
		\\
pLasso
& 4.21 (0.54) & 29.9  (0.85) & 8.08 (1.71)
& 0.44 (0.06) & 0.58 (0.03) & 0.94 (0.10)
		\\
pHS
& 0.79 (0.40) & 2.76 (0.79) & 1.35 (0.28)
& 0.99 (0.03) & 0.02 (0.06) & 0.01 (0.05)
		\\
		\hline
 \multicolumn{7}{ |c| }{ Setting VI, $n = 500$ }
\\ 
\hline
HS 
& 45.2  (3.77) & \_ & 4.67 (1.23)
& 0.99 (0.02) & 0.01 (0.05) & 0.01 (0.04)
		\\
Lasso
& 42.7  (3.67) & \_ & 4.73 (1.23)
& 0.71 (0.16) & 0.39 (0.16) & 0.49 (0.27)
		\\
pLasso
& 4.24 (0.36) & 29.6  (0.55) & 8.01 (1.70)
& 0.55 (0.10) & 0.52 (0.06) & 0.75 (0.17)
		\\
pHS
& 0.31 (0.19) & 1.00 (0.31) & 1.15 (0.21)
& 0.99 (0.03) & 0.02 (0.06) & 0.02 (0.05)
		\\
		\hline
 \multicolumn{7}{ |c| }{ Setting VI, $n = 1000$ }
\\ 
\hline
HS 
& 45.0  (2.33) & \_ & 4.54 (1.01)
& 0.99 (0.03) & 0.02 (0.05) & 0.01 (0.05)
		\\
Lasso
& 43.3  (2.38) & \_ & 4.57 (1.01)
& 0.71 (0.17) & 0.38 (0.15) & 0.48 (0.28)
		\\
pLasso
& 4.19 (0.26) & 29.6  (0.39) & 7.62 (1.70)
& 0.71 (0.12) & 0.40 (0.11) & 0.48 (0.21)
		\\
pHS
& 0.14 (0.06) & 0.48 (0.12) & 1.03 (0.20)
& 0.99 (0.03) & 0.01 (0.05) & 0.01 (0.04)
		\\
\hline	\hline	
\end{tabular}
\label{tb_seting_5_6}
\end{table}

\paragraph*{Result in variable selection}
We observe that both the Horseshoe and pliable Horseshoe methods achieve the best performance in variable selection, with the pliable Horseshoe showing a slight advantage when the sample size is relatively small ($n = 200$). However, when considering estimation and prediction errors, it is evident that the standard Horseshoe is not well-suited for capturing interaction models. In contrast, the simple lasso and pliable lasso are less effective for variable selection. The pliable lasso shows improvement over the simple lasso only when the sample size is large (e.g., $n = 1000$).

Figure \ref{fig:posterior_example1256} shows that although the Horseshoe method may correctly identify the right variable, its estimated value is skewed because it fails to account for interaction.

Trace plots and autocorrelation function (ACF) plots, presented in Figure \ref{fig_traceACF_plot} in Appendix \ref{sc_tracplot_appendix}, are provided to evaluate the convergence behavior and sampling efficiency of the Gibbs sampler.

\subsubsection{Results in high-dimensional settings}

We further evaluate the performance of our proposed methods against competing approaches in high-dimensional settings. Specifically, we focus on Setting I. While all methods performed well in low-dimensional cases, we now consider scenarios with $p = 120, n = 100$ and $p = 250, n = 200$, keeping the rest of the setup unchanged. The results, averaged over 100 simulations, are presented in Table \ref{tb_high_diems}.

As shown in Table \ref{tb_high_diems}, our proposed method (pliable Horseshoe) consistently outperforms all competitors across estimation, prediction, and variable selection accuracy. In particular, although the pliable lasso achieves lower prediction error than the standard Horseshoe or lasso, its error remains noticeably higher than that of the pliable Horseshoe. Similarly, the estimation error for our method is substantially smaller. While the pliable lasso shows improvements over the basic Horseshoe and lasso, it still lags behind our approach. For variable selection, the pliable Horseshoe achieves the best performance, followed by the standard Horseshoe. In contrast, the pliable lasso performs poorly in this aspect and appears to require larger sample sizes to be effective.

\begin{table}[!h]
\centering
	\caption{Simulation results in high dimensional settings. HS: Horseshoe; pLasso: Pliable lasso; pHS: pliable Horseshoe. Note that for HS and Lasso we do not have estimation for $\bm{\theta} $.}
	\begin{tabular}{ | l | ccc  ccc | }
		\hline \hline
Method  & ${\rm Est.} (\beta) $
& ${\rm Est.} (\bm{\theta}) $
& {\rm Pred} 
&  Accuracy
& FDR
& FPR
		\\
		\hline
 \multicolumn{7}{ |c | }{ $ n = 100 , \, p = 120 $ }
\\ 
\hline
HS 
& 11.4  (3.73) & \_ & 63.8 (20.5)
& 0.97 (0.01) & 0.00 (0.00) & 0.00 (0.00)
		\\
Lasso
& 11.4 (4.23) & \_ & 64.5 (21.7)
& 0.94 (0.06) & 0.43 (0.34) & 0.04 (0.07)
		\\
pLasso 
& 2.31 (1.40) & 4.64 (3.96) & 8.33 (5.99)
& 0.58 (0.04) & 0.93 (0.01) & 0.43 (0.05)
		\\
pHS
& 0.25 (0.98) & 0.58 (1.24) & 1.79 (1.88)
& 1.00 (0.00) & 0.01 (0.07) & 0.00 (0.00)
		\\
		\hline
 \multicolumn{7}{ |c | }{ $ n = 200 , \, p = 250 $ }
\\ 
\hline
HS 
& 7.54 (4.48) & \_ & 61.4 (22.6)
& 0.99 (0.00) & 0.01 (0.10) & 0.00 (0.00)
		\\
Lasso
& 7.17 (3.53) & \_ & 61.3 (22.4)
& 0.97 (0.03) & 0.55 (0.25) & 0.03 (0.03)
		\\
pLasso
& 0.66 (0.34) & 0.96 (0.56) & 2.77 (1.14)
& 0.69 (0.03) & 0.95 (0.00) & 0.32 (0.03)
		\\
pHS
& 0.09 (0.41) & 0.17 (0.07) & 1.30 (0.63)
& 1.00 (0.00) & 0.02 (0.06) & 0.00 (0.00)
		\\
\hline	\hline	
\end{tabular}
\label{tb_high_diems}
\end{table}

\begin{table}[!h]
\centering
\caption{Simulation results in Setting I with missing data, $ n = 200 , \, p = 10, q=4 $.}
\begin{tabular}{ | c | ccc  ccc | }
		\hline \hline
\% of missing  & ${\rm Est.} (\beta) $
& ${\rm Est.} (\bm{\theta}) $
& {\rm Pred} 
&  Accuracy
& FDR
& FPR
		\\
		\hline
10 
& 0.03 0.02 & 0.11 0.04 & 1.16 0.22
& 1.00 0.00 & 0.00 (0.00) & 0.00 (0.00)
		\\
30 
& 0.04 0.02 & 0.16 0.06 & 1.21 0.25
& 1.00 0.00 & 0.00 (0.00) & 0.00 (0.00)
\\
50
& 0.07 0.06 & 0.24 0.11 & 1.35 0.31
& 1.00 0.00 & 0.00 (0.00) & 0.00 (0.00)
\\
70
& 0.22 0.82 & 0.78 2.10 & 2.07 2.99
& 0.99 0.04 & 0.00 0.02 & 0.00 0.02
		\\
\hline	\hline	
\end{tabular}
\label{tb_simu_missingdata}
\end{table}

\subsubsection*{Result in linear model without interaction}

As we have mentioned in Section \ref{sc_model_method} that the interaction model in \eqref{eq_linear_predictor} is a generalized for linear model. We now present simulation results to assess our approach's effectiveness when applied to a standard linear model. We focus on Settings I and III, with all interaction parameters fixed at zero $\theta_0 = 0, \theta = 0$. The results from 100 simulated runs, detailed in Table \ref{tb_seting_linear_model}, indicate that all tested methods produce comparable estimation and prediction errors. Nevertheless, when it comes to selecting the correct variables, the Bayesian methods—specifically the Horseshoe and pliable Horseshoe techniques—demonstrated superior performance over their frequentist counterparts.

\subsubsection{Results with missing data}

We now present simulations in which the response variable contains missing values. Specifically, we focus on Setting I with $n = 200$, $p = 10$, and $q = 4$, while keeping all other configurations unchanged. After generating the response $y$, we randomly remove $10\%$, $30\%$, $50\%$, and $70\%$ of its values. Each experiment is repeated 100 times, and the average performance along with the corresponding standard deviation is summarized in Table \ref{tb_simu_missingdata}. The findings demonstrate that our proposed method, the pliable Horseshoe, effectively handles missing responses across estimation, prediction, and variable selection tasks. Nonetheless, as the proportion of missing data increases--for instance, at $70\%$--the performance of the method is noticeably affected.

\begin{table}[!ht]
\centering
	\caption{Simulation results in linear model without interaction, we set $\theta_0 = 0, \theta = 0$. HS: Horseshoe; pLasso: Pliable lasso; pHS: pliable Horseshoe. Note that for HS and Lasso we do not have estimation for $\bm{\theta} $.}
	\begin{tabular}{ | l | ccc  ccc | }
		\hline \hline
Method  & ${\rm Est.} (\beta) $
& ${\rm Est.} (\bm{\theta}) $
& {\rm Pred} 
&  Accuracy
& FDR
& FPR
		\\
		\hline
 \multicolumn{7}{ |c| }{ $ n=200, \rho_X = 0 $ }
\\ 
\hline
HS 
& 0.04 (0.02) & \_ &  1.05 (0.23)
& 0.99 (0.04) & 0.03 (0.07) & 0.02 (0.07)
		\\
Lasso
& 0.04 (0.03) & \_ &  1.06 (0.23)
& 0.68 (0.16) & 0.42 (0.14) & 0.54 (0.26)
		\\
pLasso
&  0.09 (0.05) & 0.05 (0.02) & 1.09 (0.19) 
& 0.66 (0.11) & 0.45 (0.08) & 0.58 (0.18)
		\\
pHS
& 0.05 (0.03) & 0.21 (0.07) &  1.30 (0.30)
& 1.00 (0.00) & 0.00 (0.00) & 0.00 (0.00)
		\\
		\hline
 \multicolumn{7}{ |c| }{ $ n=500, \rho_X = 0 $ }
\\ 
\hline
HS 
& 0.02 (0.01) & \_ & 1.05 (0.34)
& 0.99 (0.03) & 0.02 (0.07) & 0.02 (0.06)
		\\
Lasso
& 0.02 (0.01) & \_ &  1.05 (0.34)
& 0.65 (0.17) & 0.44 (0.15) & 0.59 (0.28)
		\\
pLasso
& 0.04 (0.02) & 0.01 (0.00) & 1.02 (0.21) 
& 0.86 (0.11) & 0.23 (0.16) & 0.23 (0.19)
		\\
pHS
&  0.02 (0.01) & 0.07 (0.02) &  1.13 (0.22)
& 1.00 (0.00) & 0.00 (0.00) & 0.00 (0.00)
		\\
		\hline
 \multicolumn{7}{ |c| }{ $ n=1000, \rho_X = 0 $ }
\\ 
\hline
HS 
& 0.01 (0.00) & \_ &  1.03 (0.21)
& 0.99 (0.03) & 0.03 (0.07) & 0.02 (0.06)
		\\
Lasso
& 0.01 (0.00) & \_ &  1.03 (0.21)
& 0.69 (0.17) & 0.40 (0.16) & 0.51 (0.28)
		\\
pLasso
& 0.02 (0.01) & 0.00 (0.00) & 1.03 (0.20)
& 0.94 (0.07) & 0.11 (0.12) & 0.10 (0.11)
		\\
pHS
& 0.01 (0.00) & 0.03 (0.01) &  1.06 (0.22)
& 1.00 (0.00) & 0.00 (0.00) & 0.00 (0.00)
		\\
  \hline  \hline
 \multicolumn{7}{ |c | }{ $ \rho_X = 0.5  , n=200 $ }
\\ 
\hline
HS 
& 0.06 (0.03) & \_ & 1.02 (0.20)
& 1.00 (0.02) & 0.01 (0.04) & 0.01 (0.03)
		\\
Lasso
& 0.08 (0.04) & \_ & 1.02 (0.21)
& 0.66 (0.17) & 0.42 (0.16) & 0.56 (0.29)
		\\
pLasso
& 0.11 (0.06) & 0.04 (0.02) & 1.10 (0.20)
& 0.48 (0.08) & 0.56 (0.04) & 0.87 (0.13)
		\\
pHS
& 0.08 (0.04) & 0.16 (0.10) & 1.21 (0.27)
& 1.00 (0.01) & 0.00 (0.02) & 0.00 (0.02)
		\\
		\hline
 \multicolumn{7}{ |c | }{ $ \rho_X = 0.5  , n=500  $ }
\\ 
\hline
HS 
& 0.02 (0.01) & \_ & 0.98 (0.18)
& 0.99 (0.03) & 0.02 (0.06) & 0.02 (0.06)
		\\
Lasso
& 0.03 (0.02) & \_ & 0.98 (0.19)
& 0.72 (0.15) & 0.38 (0.14) & 0.46 (0.26)
		\\
pLasso 
& 0.05 (0.03) & 0.01 (0.00) & 1.01 (0.19)
& 0.86 (0.09) & 0.24 (0.13) & 0.23 (0.15)
		\\
pHS
& 0.03 (0.01) & 0.10 (0.04) & 1.06 (0.22)
& 1.00 (0.01) & 0.00 (0.03) & 0.00 (0.02)
		\\
		\hline
 \multicolumn{7}{ |c | }{ $ \rho_X = 0.5  , n=1000  $ }
\\ 
\hline
HS 
& 0.01 (0.01) & \_ & 1.04 (0.21)
& 0.98 (0.04) & 0.03 (0.08) & 0.02 (0.07)
		\\
Lasso
& 0.01 (0.01) & \_ & 1.04 (0.21)
& 0.65 (0.16) & 0.44 (0.13) & 0.59 (0.26)
		\\
pLasso
& 0.03 (0.02) & 0.00 (0.00) & 1.08 (0.21)
& 0.93 (0.09) & 0.13 (0.14) & 0.12 (0.15)
		\\
pHS
& 0.01 (0.01) & 0.05 (0.02) & 1.07 (0.21)
& 1.00 (0.01) & 0.00 (0.00) & 0.00 (0.00)
		\\
\hline	\hline	
\end{tabular}
\label{tb_seting_linear_model}
\end{table}

\section{Application}
\label{sc_real_data}
We analyze a real dataset derived from the OASIS Brain Data project \citep{marcus2007open}. The considered dataset consists of a subset of observations from OASIS, as available in the \texttt{R} package \texttt{sail} \citep{bhatnagar2023sparse}, with additional noise variables introduced to increase the number of predictors. In total, the dataset contains measurements from 136 patients, each characterized by demographic, cognitive, and neuroimaging features.

The available covariates include patient age (Age), years of education (EDUC), Mini-Mental State Examination score (MMSE), estimated total intracranial volume (eTIV), normalized whole brain volume (nWBV), and the atlas scaling factor (ASF). These variables provide a combination of demographic and structural brain measures that are commonly studied in the context of cognitive decline and dementia research.

The outcome variable of primary interest is a continuous measure of right hippocampal volume ($y$), represented as a numeric vector of length 136. The modifying/enviromental variable is a binary dementia status indicator ($e$), with 0 corresponding to non-demented patients and 1 corresponding to demented patients. Together, these variables allow for the investigation of both structural changes in the brain and their association with clinical dementia status.

Table~\ref{tb_realdata_covariate} presents the covariates selected under the different methods. 
The variable nWBV is consistently identified across all approaches, with effect sizes of approximately 12.9 (Horseshoe), 10.7 (Lasso), 11.8 (pliable Lasso), and 12.5 (pliable Horseshoe). 
For the Bayesian procedures, we also report 95\% credible intervals. Notably, the interval for nWBV under the pliable Horseshoe is wider than that obtained with the standard Horseshoe.

In terms of interaction effects, the pliable Lasso estimates only one nonzero interaction, $\theta_{\mathrm{nWBV}} = -0.829$. In contrast, the pliable Horseshoe produces estimates $\theta_{\mathrm{nWBV}} = -3.521$ and $\theta_{\mathrm{ASF}} = 1.978$. However, neither of these interactions is deemed significant, as the corresponding 90\% credible intervals include zero.

\begin{table}[!h]
\centering
\caption{Output for real data. HS: Horseshoe; pLasso: Pliable lasso; pHS: pliable Horseshoe.}
	\begin{tabular}{  l | c | c | c  }
		\hline \hline
Method  
& selected covariates
& estimated effects
& credible interval
\\ 
\hline
HS 
& MMSE & \;\;\,0.066 & (0.014;\;\; 0.116)
		\\
& nWBV & 12.882 & (7.809; 17.495)
\\
\hline
Lasso
& EDUC     &     \;\;\,0.035
\\
& MMSE     &     \;\;\,0.052
\\
& eTIV     &      \;\;\,0.002
\\
& nWBV    &     10.763
		\\ \hline
pLasso 
& eTIV   & \;\;\,0.002
\\
& nWBV  & 11.832
		\\ \hline
pHS
& nWBV & $12.513$  &  (4.899;\;\; 21.767)
\\
& ASF & $ -5.064$ &  (-8.980;\;\; -2.082)
		\\
\hline	\hline	
\end{tabular}
\label{tb_realdata_covariate}
\end{table}

We randomly select 26 observations as the test set and use the remaining 110 observations for training. This process is repeated 100 times, and the resulting prediction errors are summarized in Table~\ref{tb_realdata}. The results indicate that both the pliable lasso and the pliable Horseshoe achieve superior predictive performance, with the pliable lasso performing best overall, while the standard Horseshoe yields the poorest accuracy.

\begin{table}[!h]
	\centering
	\caption{Mean (and standard deviation)  prediction errors for the real data. HS: Horseshoe; pLasso: Pliable lasso; pHS: pliable Horseshoe.}
	\begin{tabular}{  l | cc cc }
		\hline \hline
 & HS & Lasso & pLasso & pHS 	
 \\ 		\hline
Pred 
& 0.66 (0.24)
& 0.59 (0.20) 
& 0.54 (0.17)
& 0.56 (0.20)
\\
		\hline
		\hline
	\end{tabular}
	\label{tb_realdata}
\end{table}

\section{Discussion and conclusion}
\label{sc_conclustion}

In this paper, we propose a Bayesian pliable lasso for high-dimensional linear regression with predictor–modifier interactions. Building on the frequentist pliable lasso, our framework employs global–local shrinkage priors, specifically the horseshoe, to enforce sparsity in both main and interaction effects while naturally imposing the strong heredity constraint. This Bayesian formulation facilitates posterior inference, and uncertainty quantification, extending the interpretability and stability of the original method.

We develop an efficient Gibbs sampler exploiting conditional conjugacy through a reparametrized horseshoe prior, enabling scalable posterior computation. The approach is further extended to settings with missing responses, where unobserved outcomes are treated as latent variables within the sampling scheme. Simulation studies confirm that the method accurately recovers sparse interaction structures and provides calibrated uncertainty estimates, while applications to neuroimaging and clinical studies of dementia demonstrate its practical utility in complex biomedical data.

The method’s interpretability, ensured by the heredity constraint, is particularly valuable in scientific domains where reliable identification of effect–modifier relationships is essential. While the Gibbs sampler provides tractable inference in moderate  dimensions, further work is warranted to improve scalability for very large datasets, for example through variational inference. Extensions, for example, to survival and nonparametric models would also broaden the framework’s relevance.

\subsection*{Acknowledgments}
The views, results, and opinions presented in this paper are solely those of the author and do not, in any form, represent those of the Norwegian Institute of Public Health.

\subsection*{Conflicts of interest/Competing interests}
The author declares no potential conflict of interests.

\clearpage
\appendix
\section{Additional plots for the simulation}
\label{sc_tracplot_appendix}
\begin{figure}[!ht]
    \centering
    \includegraphics[width=12cm]{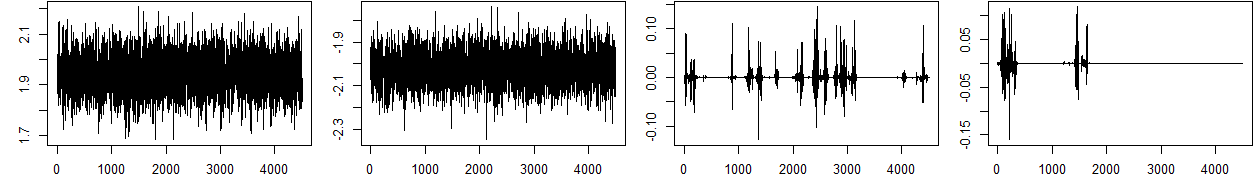}
    \includegraphics[width=12cm]{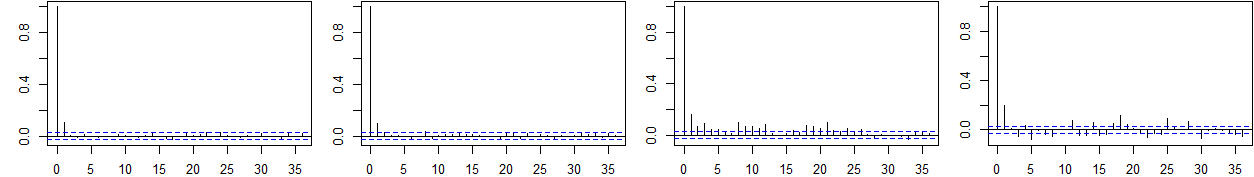}
    \caption{Trace plots and ACF plots for $ \beta_1, \beta_2 , \beta_7 , \beta_8 $ (from left to right) in Setting I with $ n =200, p = 10,q=4 $. The true values are $ \beta_1 = 2, \beta_2 = -2 , \beta_7 = \beta_8 =0 $.}
    \label{fig_traceACF_plot}
\end{figure}

\begin{figure}[!ht]
    \centering
    \includegraphics[width=12cm]{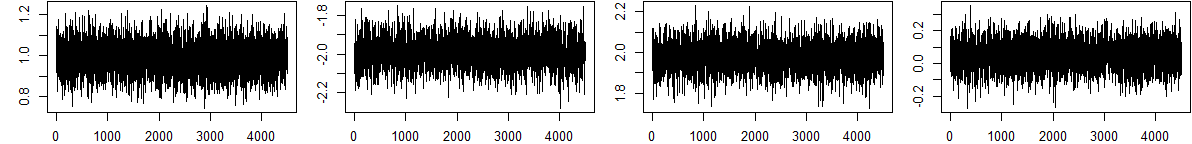}
    \includegraphics[width=12cm]{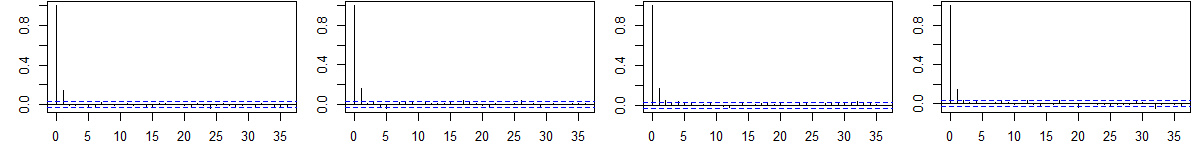}
    \caption{Trace plots and ACF plots for $ \theta_{1,2};\; \theta_{2,2} ;\; \theta_{3,2} ;\; \theta_{4,2} $ (from left to right) in Setting I with $ n =200, p = 10,q=4 $. The true values are $ \theta_{1,2} =1, \theta_{2,2}=-2 , \theta_{3,2}=2 , \theta_{4,2} =0 $.}
    \label{fig_traceACF_plot}
\end{figure}

\section{Simulations with binary response}
\label{sc_logistic}

We developed a Gibbs sampling algorithm for Bayesian inference in the logistic pliable lasso model with a hierarchical Horseshoe prior. Let $y_i \in \{0,1\}$ denote the binary response for observation $i$, $X \in \mathbb{R}^{n \times p}$ the design matrix of predictors, and $Z \in \mathbb{R}^{n \times q}$ the modifying variables. The pliable lasso augments each predictor $X_j$ with interactions with the modifying variables $Z$, leading to the linear predictor
$$
\eta_i = \beta_0 + Z_i^\top \theta_0 + \sum_{j=1}^p \Big( X_{ij}\,\beta_j + X_{ij}\, Z_i^\top \theta_j \Big),
$$
where $\beta_0$ and $\theta_0$ denote the intercept and its interactions with modifiers, $\beta_j$ is the main effect for predictor $j$, and $\theta_j$ is the vector of modifier-specific interaction effects. The Horseshoe prior is used as in the previous secctions.

The likelihood is modeled using logistic regression with Polya–Gamma (PG) data augmentation \cite{polson2013bayesian}. Specifically, conditional on latent PG variables $\omega_i \sim \text{PG}(1,\eta_i)$, the logistic likelihood admits a Gaussian form, facilitating conjugate updates.

We perform Bayesian inference for the logistic pliable-lasso model by a Gibbs sampler that combines Polya–Gamma data augmentation with a group horseshoe prior on each predictor/modifier block. Denoting the block design matrices $W_j$ (the intercept block $W_0=[1\; Z]$ and for $j\ge1$ the block $W_j=[x_j\; x_j\odot Z]$) and the block parameters $\gamma_j =(\beta_j, \theta_j^\top)^\top \in\mathbb{R}^{1+q}$, the Polya–Gamma augmentation 
$$
\omega_i\sim\mathrm{PG}(1,\eta_i)
$$ 
leads to a conditional Gaussian likelihood proportional to 
$$
\exp(\kappa^\top\eta - \tfrac12\eta^\top\Omega\eta)
$$ 
with $\kappa=y-\tfrac12$
 and $\Omega=\mathrm{diag}(\omega)$. Conditional on $\omega$ and the horseshoe scales $(\tau^2,\lambda_j^2)$ each block $\gamma_j$ has a multivariate normal full conditional
$$
\gamma_j\mid\cdot \sim \mathcal{N}\big(V_j W_j^\top(\kappa-\Omega\eta_{-j}),\; V_j\big),\qquad V_j=(W_j^\top\Omega W_j + (\tau^2\lambda_j^2)^{-1}I_{1+q})^{-1},
$$
(with the intercept block using prior precision $\sigma_0^{-2}I_{1+q}$). 
The auxiliary inverse-Gamma representation of the half-Cauchy yields closed-form inverse-Gamma posteriors for the local scales
$$
\lambda_j^2\sim\mathrm{IG}((d+1)/2,\; 1/\nu_j + \|\gamma_j\|^2/(2\tau^2)),
\quad
\nu_j\sim\mathrm{IG}(1/2,1+1/\lambda_j^2)
,
$$ 
and likewise for the global scale 
$$
\tau^2\sim\mathrm{IG}((pd+1)/2,\; 1/\xi + \tfrac12\sum_j \|\gamma_j\|^2/\lambda_j^2)
,
\quad
\xi\sim\mathrm{IG}(1/2,1+1/\tau^2)
.
$$
In practice we stabilize matrix inversions with a small ridge term and draw Polya–Gamma variates via the \texttt{BayesLogit} package. Efficient incremental updates of $\eta$ and Cholesky sampling of the $d\times d$ posteriors are used for speed.
\begin{table}[!ht]
\centering
	\caption{Simulation results in logistic regression model with $ p = 10,q = 4 $ for Setting I and II. HS: Horseshoe; pHS: pliable Horseshoe. Note that for HS and Lasso, we do not have estimation for $\bm{\theta} $.}
	\begin{tabular}{ | r | ccc  ccc | }
		\hline \hline
Method  & ${\rm Est.} (\beta) $
& ${\rm Est.} (\bm{\theta}) $
& {\rm Pred} 
&  Accuracy
& FDR
& FPR
		\\
		\hline
 \multicolumn{7}{| c | }{ Setting I, $n = 200$ }
\\ 
\hline
HS 
& 10.5  (1.36) &  & 0.32 (0.06)
& 0.82 (0.09) & 0.00 (0.03) & 0.00 (0.02)
		\\
Lasso
& 11.5 (1.41) &  & 0.32 (0.07)
&  0.76 (0.15) & 0.32 (0.18) & 0.36 (0.26)
		\\
pHS
& 1.35 (1.31) & 4.65 (3.65) & 0.13 (0.05)
& 0.99 (0.03) & 0.00 (0.00) & 0.00 (0.00)
		\\
		\hline
 \multicolumn{7}{ | c | }{  Setting I,  $n = 500$ }
\\ 
\hline
HS 
& 10.0  (0.84) &  & 0.30 (0.07)
& 0.95 (0.06) & 0.01 (0.05) & 0.01 (0.04)
		\\
Lasso
& 10.6  (0.98) &  & 0.30 (0.07)
& 0.70 (0.16) & 0.39 (0.17) & 0.50 (0.27)
		\\
pHS
& 0.41 (0.36) & 1.45 (0.93) & 0.11 (0.05)
& 1.00 (0.00) & 0.00 (0.00) & 0.00 (0.00)
		\\
		\hline
 \multicolumn{7}{ | c | }{  Setting I,  $n = 1000$ }
\\ 
\hline
HS 
& 9.71 (0.50) &  & 0.29 (0.07)
& 0.99 (0.03) & 0.01 (0.05) & 0.01 (0.04)
		\\
Lasso
& 10.2 (0.51) &  & 0.30 (0.07)
& 0.70 (0.16) & 0.39 (0.15) & 0.49 (0.27)
		\\
pHS
& 0.19 (0.22) & 0.67 (0.53) & 0.10 (0.04)
& 1.00 (0.00) & 0.00 (0.00) & 0.00 (0.00)
		\\
\hline		\hline
 \multicolumn{7}{ | c | }{  Setting II,  $n = 200$ }
\\ 
\hline
HS 
& 23.2 (25.7) &  & 0.08 (0.04)
& 0.99 (0.03) & 0.01 (0.05) & 0.01 (0.04)
		\\
Lasso
& 6.52 (5.52) &  & 0.08 (0.04)
& 0.59 (0.13) & 0.49 (0.10) & 0.69 (0.22)
		\\
pHS
& 4.10 (4.53) & 12.4 (7.77) & 0.08 (0.04)
& 0.92 (0.07) & 0.00 (0.00) & 0.00 (0.00)
		\\
		\hline
 \multicolumn{7}{ | c| }{ Setting II, $n = 500$ }
\\ 
\hline
HS 
& 14.9  (7.94) &  & 0.08 (0.04)
& 0.99 (0.03) & 0.01 (0.05) &  0.01 (0.05)
		\\
Lasso
& 7.60 (4.23) &  & 0.08 (0.04)
& 0.54 (0.11) & 0.53 (0.07) & 0.76 (0.18)
		\\
pHS
& 1.54 (1.21) & 6.80 (3.87) &  0.07 (0.04)
& 0.99 (0.03) & 0.00 (0.00) & 0.00 (0.00)
		\\
		\hline
 \multicolumn{7}{ |c| }{ Setting II, $n = 1000$ }
\\ 
\hline
HS 
& 12.7 (3.92) &  & 0.07 (0.04)
& 0.99 (0.04) & 0.03 (0.07) & 0.02 (0.06)
		\\
Lasso
& 8.55 (3.02) &  & 0.07 (0.04)
& 0.50 (0.10) & 0.55 (0.06) & 0.82 (0.16)
		\\
pHS
& 0.73 (0.69) & 4.39 (2.63) & 0.06 (0.04)
& 1.00 (0.00) & 0.00 (0.00) & 0.00 (0.00)
		\\
\hline	\hline	
\end{tabular}
\label{tb_seting_1_2_logistic}
\end{table}


\begin{thebibliography}{}
	
	\bibitem[Andrinopoulou and Rizopoulos, 2016]{andrinopoulou2016bayesian}
	Andrinopoulou, E.-R. and Rizopoulos, D. (2016).
	\newblock Bayesian shrinkage approach for a joint model of longitudinal and
	survival outcomes assuming different association structures.
	\newblock {\em Statistics in medicine}, 35(26):4813--4823.
	
	\bibitem[Asenso et~al., 2024]{asenso2024pliable}
	Asenso, T.~Q., Wang, P., and Zhang, H. (2024).
	\newblock Pliable lasso for the support vector machine.
	\newblock {\em Communications in Statistics-Simulation and Computation},
	53(2):786--798.
	
	\bibitem[Asenso et~al., 2022]{asenso2022pliable}
	Asenso, T.~Q., Zhang, H., and Liang, Y. (2022).
	\newblock Pliable lasso for the multinomial logistic regression.
	\newblock {\em Communications in Statistics-Theory and Methods},
	51(11):3596--3611.
	
	\bibitem[Bhadra et~al., 2019]{bhadra2019lasso}
	Bhadra, A., Datta, J., Polson, N.~G., and Willard, B.~T. (2019).
	\newblock Lasso meets horseshoe: A survey.
	\newblock {\em Statistical Science}, 34(3):405--427.
	
	\bibitem[Bhatnagar et~al., 2025]{bhatnagar2023sparse}
	Bhatnagar, S., Yang, Y., and Greenwood, C. (2025).
	\newblock {\em sail: Sparse Additive Interaction Learning}.
	\newblock R package version 0.1.0, commit
	f5c4b09a396c667aedc4e167b4fca5e65f5c5f51.
	
	\bibitem[Bien et~al., 2013]{bien2013lasso}
	Bien, J., Taylor, J., and Tibshirani, R. (2013).
	\newblock A lasso for hierarchical interactions.
	\newblock {\em Annals of statistics}, 41(3):1111.
	
	\bibitem[B{\"u}hlmann and Van De~Geer, 2011]{buhlmann2011statistics}
	B{\"u}hlmann, P. and Van De~Geer, S. (2011).
	\newblock {\em Statistics for high-dimensional data: methods, theory and
		applications}.
	\newblock Springer Science \& Business Media.
	
	\bibitem[Carvalho et~al., 2009]{carvalho2009handling}
	Carvalho, C.~M., Polson, N.~G., and Scott, J.~G. (2009).
	\newblock Handling sparsity via the horseshoe.
	\newblock {\em Journal of Machine Learning Research W\&CP}, 5:73--80.
	
	\bibitem[D'Alessandro et~al., 2025]{d2025integrating}
	D'Alessandro, M., Asenso, T.~Q., and Zucknick, M. (2025).
	\newblock Integrating multiple data sources with interactions in multi-omics
	using cooperative learning.
	\newblock {\em Statistics in Medicine}, 44(13-14):e70148.
	
	\bibitem[Du and Tibshirani, 2018]{du2018pliable}
	Du, W. and Tibshirani, R. (2018).
	\newblock A pliable lasso for the cox model.
	\newblock {\em arXiv preprint arXiv:1807.06770}.
	
	\bibitem[Friedman et~al., 2010]{glmnet}
	Friedman, J., Hastie, T., and Tibshirani, R. (2010).
	\newblock Regularization paths for generalized linear models via coordinate
	descent.
	\newblock {\em Journal of Statistical Software}, 33(1):1--22.
	
	\bibitem[Giraud, 2021]{giraud2021introduction}
	Giraud, C. (2021).
	\newblock {\em Introduction to high-dimensional statistics}.
	\newblock Chapman and Hall/CRC.
	
	\bibitem[Hastie et~al., 2009]{hastie2009elements}
	Hastie, T., Tibshirani, R., Friedman, J.~H., and Friedman, J.~H. (2009).
	\newblock {\em The elements of statistical learning: data mining, inference,
		and prediction}, volume~2.
	\newblock Springer.
	
	\bibitem[Hoogland et~al., 2021]{hoogland2021tutorial}
	Hoogland, J., IntHout, J., Belias, M., Rovers, M.~M., Riley, R.~D.,
	E.~Harrell~Jr, F., Moons, K.~G., Debray, T.~P., and Reitsma, J.~B. (2021).
	\newblock A tutorial on individualized treatment effect prediction from
	randomized trials with a binary endpoint.
	\newblock {\em Statistics in medicine}, 40(26):5961--5981.
	
	\bibitem[Kim et~al., 2021]{kim2021svreg}
	Kim, R., M{\"u}ller, S., and Garcia, T.~P. (2021).
	\newblock svreg: Structural varying-coefficient regression to differentiate how
	regional brain atrophy affects motor impairment for huntington disease
	severity groups.
	\newblock {\em Biometrical Journal}, 63(6):1254--1271.
	
	\bibitem[Lim and Hastie, 2015]{lim2015learning}
	Lim, M. and Hastie, T. (2015).
	\newblock Learning interactions via hierarchical group-lasso regularization.
	\newblock {\em Journal of Computational and Graphical Statistics},
	24(3):627--654.
	
	\bibitem[Mai, 2024]{mai2024concentration}
	Mai, T.~T. (2024).
	\newblock {Concentration of a Sparse Bayesian Model With Horseshoe Prior in
		Estimating High-Dimensional Precision Matrix}.
	\newblock {\em Stat}, 13(4):e70008.
	
	\bibitem[Mai, 2025a]{mai2025handling}
	Mai, T.~T. (2025a).
	\newblock {Handling bounded response in high dimensions: a Horseshoe prior
		Bayesian Beta regression approach}.
	\newblock {\em arXiv preprint arXiv:2505.22211}.
	
	\bibitem[Mai, 2025b]{mai2025hightobit}
	Mai, T.~T. (2025b).
	\newblock {High-dimensional Bayesian Tobit regression for censored response
		with Horseshoe prior}.
	\newblock {\em arXiv preprint arXiv:2505.08288}.
	
	\bibitem[Makalic and Schmidt, 2016]{makalic_simple_2016}
	Makalic, E. and Schmidt, D.~F. (2016).
	\newblock A simple sampler for the horseshoe estimator.
	\newblock {\em IEEE Signal Processing Letters}, 23(1):179--182.
	
	\bibitem[Marcus et~al., 2007]{marcus2007open}
	Marcus, D.~S., Wang, T.~H., Parker, J., Csernansky, J.~G., Morris, J.~C., and
	Buckner, R.~L. (2007).
	\newblock Open access series of imaging studies (oasis): cross-sectional mri
	data in young, middle aged, nondemented, and demented older adults.
	\newblock {\em Journal of cognitive neuroscience}, 19(9):1498--1507.
	
	\bibitem[McCullagh and Nelder, 1989]{mccullagh1989generalized}
	McCullagh, P. and Nelder, J.~A. (1989).
	\newblock {\em Generalized linear models}.
	\newblock Monographs on Statistics and Applied Probability. Chapman \& Hall,
	London, 2nd edition.
	
	\bibitem[Polson et~al., 2013]{polson2013bayesian}
	Polson, N.~G., Scott, J.~G., and Windle, J. (2013).
	\newblock Bayesian inference for logistic models using p{\'o}lya--gamma latent
	variables.
	\newblock {\em Journal of the American statistical Association},
	108(504):1339--1349.
	
	\bibitem[Sun et~al., 2024]{sun2024bhaft}
	Sun, N., Chu, J., He, Q., Wang, Y., Han, Q., Yi, N., Zhang, R., and Shen, Y.
	(2024).
	\newblock Bhaft: Bayesian heredity-constrained accelerated failure time models
	for detecting gene-environment interactions in survival analysis.
	\newblock {\em Statistics in Medicine}, 43(21):4013--4026.
	
	\bibitem[Tibshirani and Friedman, 2020]{tibshirani2020pliable}
	Tibshirani, R. and Friedman, J. (2020).
	\newblock A pliable lasso.
	\newblock {\em Journal of Computational and Graphical Statistics},
	29(1):215--225.
	
	\bibitem[{van der Pas} et~al., 2019]{horseshoe_package}
	{van der Pas}, S., Scott, J., Chakraborty, A., and Bhattacharya, A. (2019).
	\newblock {\em horseshoe: Implementation of the Horseshoe Prior}.
	\newblock R package version 0.2.0.
	
	\bibitem[van~der Pas et~al., 2017]{van2017adaptive}
	van~der Pas, S., Szab{\'o}, B., and van~der Vaart, A. (2017).
	\newblock Adaptive posterior contraction rates for the horseshoe.
	\newblock {\em Electronic Journal of Statistics}, 11:3196--3225.
	
	\bibitem[Wu et~al., 2014]{wu2014integrative}
	Wu, C., Cui, Y., and Ma, S. (2014).
	\newblock Integrative analysis of gene--environment interactions under a
	multi-response partially linear varying coefficient model.
	\newblock {\em Statistics in medicine}, 33(28):4988--4998.
	
	\bibitem[Wu et~al., 2018]{wu2018identifying}
	Wu, M., Huang, J., and Ma, S. (2018).
	\newblock Identifying gene-gene interactions using penalized tensor regression.
	\newblock {\em Statistics in medicine}, 37(4):598--610.
	
\end{thebibliography}
\end{document}